\documentclass[12pt,preprint]{aastex}
\usepackage{emulateapj5}

\slugcomment{To appear in ApJ Letters}

\newcommand\asca{{\it ASCA}}

\newcommand\chandra{{\it Chandra}}
\newcommand\rosat{{\it ROSAT}}

\newcommand\xmm{{\it XMM-Newton}}

\newcommand\s{{\rm~s}}
\newcommand\ks{{\rm~ks}}
\newcommand\kev{{\rm~keV}}
\newcommand\ev{{\rm~eV}}
\newcommand\kms{\ifmmode {\rm~km\ s}^{-1} \else ~km s$^{-1}$\fi}
\newcommand\Hunit{\ifmmode {\rm~km\ s}^{-1}\ {\rm Mpc}^{-1}
        \else ~km s$^{-1}$ Mpc$^{-1}$\fi}
\newcommand\ctssec{\ifmmode {\rm~count\ s}^{-1} \else ~count s$^{-1}$\fi}
\newcommand\ergsec{\ifmmode {\rm~erg\ s}^{-1} \else
        ~erg s$^{-1}$\fi}
\newcommand\funit{\ifmmode {\rm~erg\ s}^{-1}\;{\rm cm}^{-2} \else
        ~ergs s$^{-1}$ cm$^{-2}$\fi}
\newcommand\phflux{\ifmmode {\rm~photon\ s}^{-1}\;{\rm cm}^{-2}
        \else   ~photon s$^{-1}$ cm$^{-2}$\fi}
\newcommand\efluxA{\ifmmode {\rm~erg\ s}^{-1}\;{\rm cm}^{-2}\;{\rm
        \AA}^{-1} \else ~erg s$^{-1}$ cm$^{-2}$ \AA$^{-1}$\fi}
\newcommand\efluxHz{\ifmmode {\rm~erg\ s}^{-1}\;{\rm cm}^{-2}\;{\rm
        Hz}^{-1} \else ~erg s$^{-1}$ cm$^{-2}$ Hz$^{-1}$\fi}
\newcommand\cc{\ifmmode {\rm~cm}^{-3} \else cm$^{-3}$\fi}
\newcommand\FWHM{\ifmmode {\rm~FWHM} \else ${\rm~FWHM}$\fi}
\newcommand\Msun{\ifmmode M_{\odot} \else $M_{\odot}$\fi}
\newcommand\Lsun{\ifmmode L_{\odot} \else $L_{\odot}$\fi}
\newcommand\ltsim{\raisebox{-.5ex}{$\;\stackrel{<}{\sim}\;$}}

\newcommand\hbeta{\ifmmode {\rm H}\beta \else H$\beta$\fi}
\newcommand\Kalpha{\ifmmode {\rm K}\alpha \else K$\alpha$\fi}
\newcommand\nh{\ifmmode N_{\rm H} \else N$_{\rm H}$\fi}
\usepackage{graphicx}
\usepackage{here}

\begin{document}

\title{A  transition to a low/soft state in the ultra-luminous compact X-ray
  source Holmberg II X-1}

\author{Gulab C. Dewangan\altaffilmark{1}, Takamitsu Miyaji\altaffilmark{1}, Richard E. Griffiths\altaffilmark{1}, \& Ingo Lehmann\altaffilmark{2}} 
\altaffiltext{1}{Department of Physics, Carnegie Mellon University, 5000 Forbes Avenue, Pittsburgh, PA 15213 USA}
\altaffiltext{2}{Max-Planck-Institut für extraterrestrische Physik, Giessenbachstrasse, PF 1312, 85741 Garching, Germany}

\begin{abstract}
We present three \xmm{} observations of the ultra-luminous compact X-ray
source Holmberg II X-1 in its historical brightest and faintest states. The
source  was in its brightest state in April
2002 with an isotropic X-ray luminosity of $\sim 2\times
10^{40}{\rm~erg~s^{-1}}$  but changed to a peculiar low/soft state in
September  2002 in which the
X-ray flux dropped by a factor of $\sim 4$ and the spectrum softened. In all
cases, a soft excess component, which can be described by a simple or 
multicolor disk blackbody (MCD; $kT \sim 120-170\ev$), is 
statistically required in addition to a
power-law continuum ($\Gamma \sim 2.4 - 2.9$). Both
the spectral components became weaker and softer in the low/soft state, 
however, the dramatic variability is seen in the power-law component. This
spectral transition is opposite to the `canonical' high/soft -- low/hard
transitions seen in many Galactic black hole binaries. 
There is possible contribution from an optically thin thermal
plasma. When this component is taken into account, the spectral transition
appears to be  normal -- a drop of the
power-law flux and slightly softer blackbody component in the low state.
\end{abstract}

\keywords{accretion disks  --- stars: individual (Holmberg II X-1) --- X-rays: stars}

\section{Introduction}
Ultra-luminous X-ray sources (ULXs) are the compact, off-nuclear galactic
X-ray sources with isotropic luminosity exceeding the Eddington limit for
a stellar mass ($\ltsim 20M\sun$) black hole (BH). Isotropic emission from
intermediate mass ($\sim 20-10^5M\sun$) BHs and anisotropic emission from
stellar  mass BHs are both
attractive interpretation for the nature of ULXs, however, there is no
definite proof for either of the two scenarios (Miller \& Colbert 2003).
Several observations suggest that ULXs may
be similar to the X-ray binaries. Discovery of orbital modulations from
several ULXs (Bauer et al. 2001; Sugiho et al. 2001) implies binary nature.  
The \asca{} X-ray spectra of ULXs have been
described as the emission from optically thick accretion disks 
 (Makishima et al. 2000 and references therein),
however, the inferred disk temperatures, in the range of $\sim 1-2\kev$, are
too high for BH masses
implied by their X-ray luminosity.  
Recent observations with \chandra{} and \xmm{} suggest lower
temperatures in the range of $\sim 100 -500\ev$ (see review by Miller \&
Colbert 2003 and references therein).
Observations of spectral transitions between low (hard) and high (soft) state
from two ULXs in IC~342 (Kubota et al. 2001) further demonstrate their
similarity with the Galactic BH binaries. These observational facts suggest that ULXs may be
more massive or scaled up versions of X-ray binaries. Recently, Ebisawa et
al. (2003) argued that the high temperatures inferred from the X-ray spectra of
some ULXs is high even for an accreting Kerr BH, and
they explained the X-ray spectra in terms of a slim disk 
( model without requiring
intermediate mass BHs.

 Holmberg II  X-1
(hereafter HoII X-1) is nearby ($3.39{\rm~Mpc}$; Karachentsev et al. 2002) and
one of the brightest ULX in the sky. Its X-ray luminosity ($L_X \sim
10^{40}{\rm~erg~s^{-1}}$) corresponds to the Eddington luminosity of an
$80M_{\sun}$ BH. It resides near the edge of a dwarf star-forming
  galaxy and is embedded in an H~\small{II} region. Also a compact HeII 
  emission-line cloud has been discovered at the position of HoII X-1, indicating
  that this ULX is actually photo-ionizing the surrounding gas (Pakull \&
  Mirioni 2001; Lehmann et al. 2004). The \rosat{} HRI and PSPC observations
  of HoII X-1 revealed a point-like, variable source on scales of days and
  years (Zezas et al. 1999). The \chandra{} ACIS-S observations also revealed
  point-like X-ray source and an indication for a weak extended component
  (Lehmann  et al 2004). Miyaji, Lehmann, \& Hasinger
  (2001) presented the joint \rosat{} and \asca{} spectrum, which is well
  described 
  by a power law ($\Gamma_X \sim 1.9$) and a soft excess component modeled
  either as an MCD ($kT \sim 170 \ev$) or a mekal plasma 
($kT\sim 300\ev$). In this paper, we present puzzling X-ray spectral variability  of this well
known ULX Holmberg II X-1 observed with \xmm{}.

\section{Observation and Data Reduction}
\xmm{} observed Ho II X-1 three times on 2002 April 10, 16, and
September 18 for $12.6$, $13.8$, and $6.9\ks$, respectively.  
The EPIC PN \citep{Struderetal01} and MOS \citep{Turneretal01} cameras were 
operated in full frame mode using the thin filter. 
%
%
The raw events were processed and filtered using the most recent
updated calibration database and analysis software ({\tt SAS v5.4.1})
available in December 2003. Cleaning of the flaring particle background 
resulted in `good' exposure times of $4.66$, $3.95$, and $4.25\ks$ for
the  EPIC PN observations of 10 April, 16 April, and 18 September 2002,
respectively.
 Events in the bad pixels and those
adjacent pixels were discarded.  Only events with pattern $0-4$ (single and
double) for the PN and $0-12$ for the MOS were selected.  
 
\section{Analysis \& Results}
We extracted full band light curves for HoII X-1 from the PN data using 
circular regions of radii $40\arcsec$ centered at the source position. 
We also extracted the corresponding background light curves from source free
regions with the same bin sizes and exposure requirement. The background
corrected PN light curves of HoII X-1 are shown in Figure~1 for the three 
\xmm{} observations. Small amplitude ($\ltsim 20\%$) fluctuations are seen
in each of the three observations. A constant count rate fit to each 
light curve resulted in $\chi^2$/dof $= 62.0/51$, $56.5/46$, and $40.3/46$ for
the observations of 10, 16 April 2002, and 18 Sep. 2002.  This suggests that X-ray emission was variable during the
first two observations of April 2002. The mean flux level remained similar in
these two observations.  However, X-ray intensity from HoII X-1 dropped 
by a factor of $\sim 3-4$ during the third observations of September 2002.
For comparison we show the long-term light curve of HoII X-1 in Fig.~\ref{f1},
derived for the common energy band of \rosat{}, \asca{}, and \xmm{}.
The \rosat{} and \asca{} fluxes were taken from Miyaji et al. (2001). It is
clear that the source was in its brightest and faintest flux levels during the
\xmm{} observations of 10 April 2002 and 18 Sep. 2002, respectively.
  
Source spectra and the associated background spectra were extracted using
similar extraction regions described above. 
The source spectra were grouped to
a minimum of 20 counts and 15 counts per channel for the 2002 April
and September observations, respectively. The spectra  were
analyzed using {\tt XSPEC 11.3}.  All the errors quoted 
below were calculated at $90\%$ confidence level for one interesting parameter.
We found generally 
good agreement
between MOS and PN cameras in the $0.3-10\kev$ bands.  Therefore we present 
the spectral results obtained by fitting the same model jointly to the PN 
and MOS data while leaving the relative normalizations to vary.
We fitted an absorbed power-law model to the spectra of 
10 April 2002, this
resulted in $\chi^2=933.4$ for 758 degrees of freedom (dof). The best-fit
absorption column is $\nh \sim 1.9\times 10^{21}{\rm~cm^{-2}}$, much higher 
than the Galactic column of $3.4\times 10^{20}{\rm~cm^{-2}}$ (Dickey \& Lockman
  1990) along the direction of Ho~II X-1. The excess column, inferred from an
  additional absorption model, is $1.6_{-0.1}^{+0.2}\times
  10^{21}{\rm~cm^{-2}}$. Addition of blackbody (BB) model improves the fit
  significantly (model 1; $\Delta \chi^2 = -62.5$ for two additional
  parameters).  The
  BB temperature is $kT \sim 140\ev$ and power-law photon index is
  $\Gamma_X \sim 2.6$. The EPIC spectral data of 10 April 2002 and the 
best-fit model is shown
  in Figure~\ref{f2}. Replacing the simple BB component by an MCD (model 2), appropriate for a standard thin accretion disk,
  resulted in a slightly higher temperature. The best-fit model parameters are
  listed in Table~\ref{tab1}. We followed the above steps to model the EPIC spectra
  of HoII X-1 obtained on 16 April 2002 and 18 Sep. 2002 and found that these
  spectra too are well described by the absorbed power law and BB model
  with the addition of the later component improving the fit significantly
  ($\Delta \chi^2 = -61.8$, and $-26.7$ for the spectra of April 10, and 
Sep. 18, respectively). The best-fit parameters for the three spectra are
listed in Table~\ref{tab1}. The X-ray spectrum of 18 Sep. 2002 is weakest and steepest 
among the three spectra. The power-law normalization decreased by a factor of
about four in five months. However, X-ray spectrum of HoII X-1 is not the
flattest at its brightest phase i.e., there is no evidence for correlation between the
spectral slope and the flux. 

To illustrate the flux and  spectral variability of HoII X-1, we compared 
each of the observed spectra with the best-fit spectral model to the spectrum
of 10 April 2002. Figure~\ref{f3} shows ratios of the individual spectra and
the best-fit model spectrum of 10 April 2002.
It is clear that the spectrum of HoII X-1 did not change appreciably during
the two observations in April 2002, but became drastically weaker and steeper 
in Sep. 2002.     

The soft X-ray line-like features in the low state spectrum are
suggestive of an optically-thin emission. Unfortunately, RGS data is too poor to
constrain any line emission.
The line-like feature at $2.5\kev$, seen in the 
low state spectrum only (see Fig.~\ref{f3}), is detected at
$2.8\sigma$ level or $98\%$ confidence level based on a maximum likelihood
ratio (MLR) test ($\Delta \chi^2 = -7.8$ for two additional parameters) using the
model~1 (see Table~1) as the continuum and a Gaussian line. This feature, also
seen in the \asca{} spectra ($\Delta \chi^2 = -7.1$ for GIS+SIS data using the
same continuum model), is likely the S~XV K$\alpha$
emission line. There may be some contribution from the radiative recombination 
continuum (RRC) due to Si~\small{XIII}.  To investigate the effect of the presence of thin
thermal emission on the spectral variability, we created a multi-component model consisting
of a mekal plasma, a BB, and a power-law modified by the
Galactic as well excess absorption. We fitted this model jointly to the PN
spectra corresponding to the brightest flux state of 10 April 2002 and faintest
flux state of 18 September 2002. The parameters of the BB and the power-law components
for the two data sets 
were varied independently, while the parameters of the mekal component
for low state spectrum were tied to the corresponding parameters for the high
state spectrum. This model resulted in a good fit ($\chi^2 = 504.2$ for 499
dof).  The unfolded spectra along with the models for the low and high states
are plotted in Figure~\ref{f4}. The best-fit parameters are: 
$kT_{mekal} = 0.90_{-0.12}^{+0.22}\kev$, emission
measure $EM = 9.3\times 10^{60}{\rm~cm^{-3}}$,
$kT_{BB} (high) = 140_{-10}^{+7}\ev$, $\Gamma (high) = 2.6_{-0.1}^{+0.1}$,
$kT_{BB}(low) = 106_{-13}^{+7}\ev$, $\Gamma (low) = 2.7_{-0.1}^{+0.2}$. The power-law normalization in the low
state is a factor of $5.2$ lower than that in the high state ($2.4\times
10^{-3}{\rm~photons~cm^{-2}~s^{-1}~keV^{-1}}$ at $1\kev$). The two photon
indices are similar within errors.   Thus, the softer
spectrum in the low state can be explained as the drop of the flux in the
power-law component without significantly changing its slope and a slightly
softer BB component. The exclusion of the mekal component worsened the
fit, resulting in  $\chi^2 = 512.6$
for 501 dof. Thus, the presence of the mekal component is significant at a
confidence level of $98.5\%$ based on the MLR test. In the low state alone, the presence of the mekal
component is  significant at $99.3\%$ level ($\Delta \chi^2 = -10.1$ for two
additional parameters).

\section{Discussion}
\begin{figure}
   \centering
   \includegraphics[height=7cm]{f1a.ps}
  \includegraphics[height=7cm]{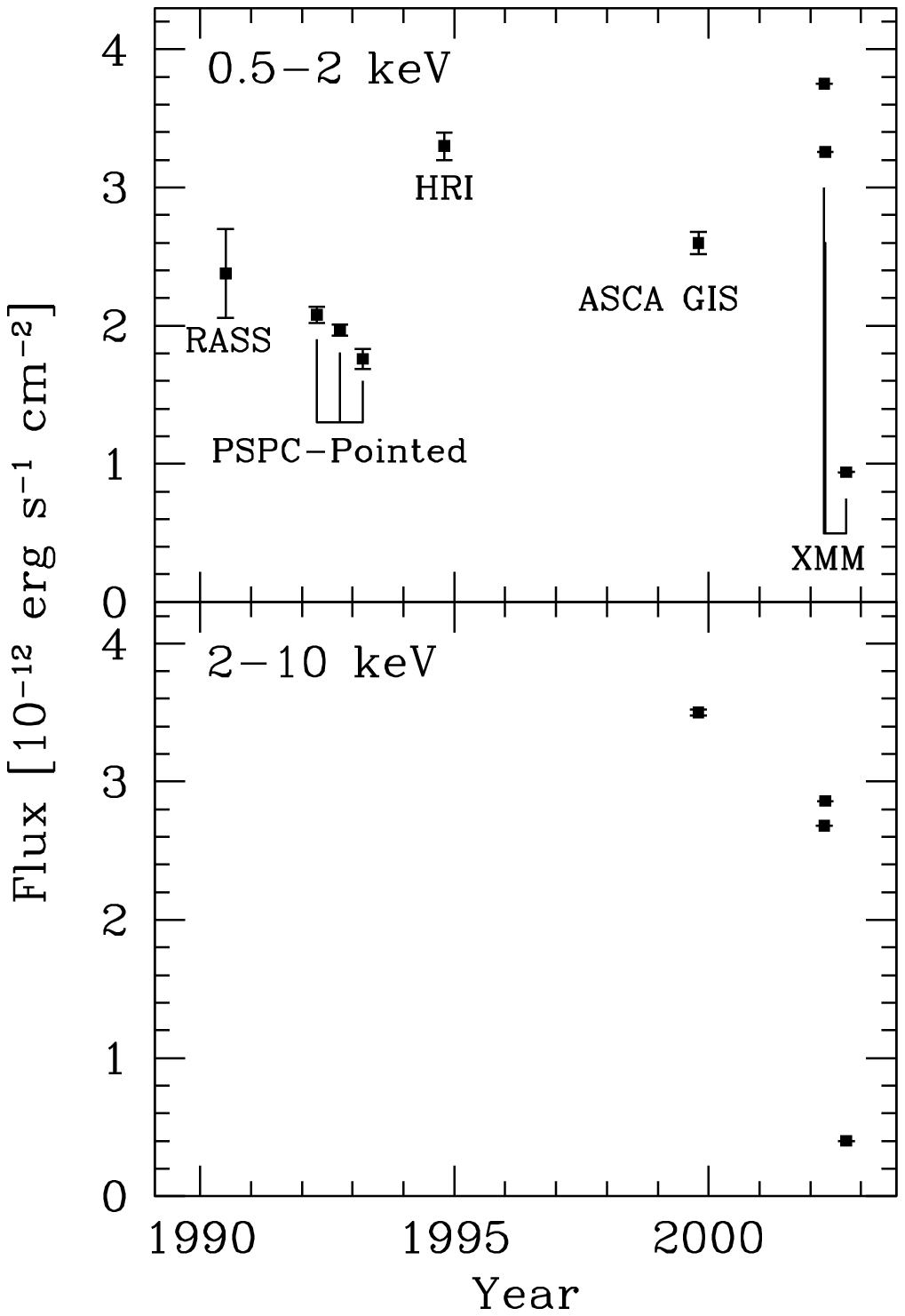}
\caption{{\it Left:} EPIC PN light curves of Holmberg II X-1 obtained
  from the three observations (a) 10 April 2002, (b) 16 April 2002, and (c) 18
  September 2002. The light curves have been corrected
  for the background contribution and are shown with time bins of $100\s$.
{\it Right:} Long-term light curve of HoII X-1. \rosat{} and \asca{} fluxes
  were taken from Miyaji et al. (2001). } 
\label{f1}
\end{figure}

\begin{figure}
 \centering
\includegraphics[width=6cm,angle=-90]{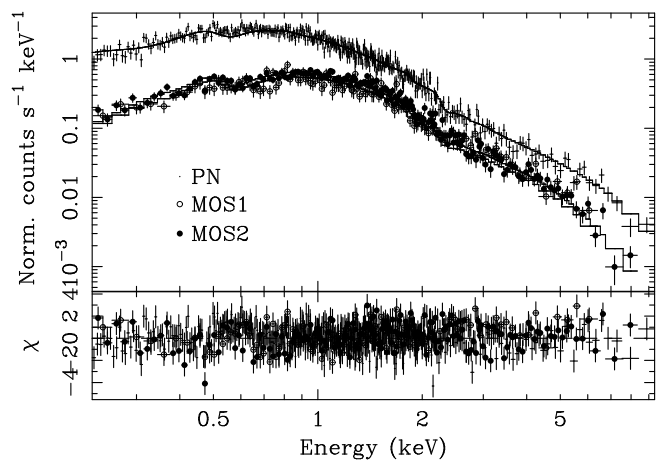}
\caption{The observed spectra of Holmberg II X-1 and the
  best-fit model consisting of a power law and a BB modified by the
  Galactic as well as the intrinsic absorption. The bottom panel shows the
  deviations of the observed data points from the best-fit model in units of $\sigma$.}
\label{f2}
\end{figure}

The three \xmm{} observations of 2002 have caught HoII X-1 in its historical
brightest 
and faintest flux levels ($3.7\times10^{-12}$ and
$9.4\times10^{-13}{\rm~erg~cm^{-2}~s^{-1}}$ in the $0.5-2\kev$ band,
respectively) ever observed with an X-ray satellite (see Fig.~\ref{f1}). 
Our spectral analysis of the  three \xmm{} observations of HoII X-1 confirm
earlier \rosat{} and \asca{} results that the X-ray spectrum of HoII X-1
consists of three spectral components - intrinsic absorption, a soft excess, 
and a power-law component.
The intrinsic absorption column densities and temperatures ($150-190\ev$) of 
the soft component inferred from the MCD fits to three observations are similar to that obtained from the
joint \rosat{} and \asca{} spectral analysis (Miyaji et al. 2001). However,
the power-law photon indices are steeper ($\Delta \Gamma \sim 0.5 - 1.0$)
during the \xmm{} observations than that during the \asca{} observations.   

%
A soft X-ray excess component, above a power-law continuum, is statistically
required in all three observations and is well described by a simple BB
or an MCD model. The best-fit temperatures are in the
range 
$100-180\ev$. Similar cool thermal components have been recently observed by
\chandra{} and \xmm{} from a number of bright ULXs e.g., M81 X-9 (Miller,
Fabian, \& Miller 2003a), NGC~4038/4039 X-37 (Miller et al. 2003b), and NGC~5048
X-1 (Karret et al. 2003). 
If the soft
X-ray  excess emission is attributed to the disk
emission, as thought to be the case in BH X-ray binaries, the accretion disk of
HoII X-1 must be cool. In this scenario, the BB temperature provides a
BH mass  
\begin{equation}
M \simeq 10 (kT/1.2\kev)^{-4}
(\dot{M}/\dot{M}_{Edd}) M_{\sun}
\label{a}
\end{equation}
(see e.g. Makishima et al. 2000). 
 If we assume that a fraction $\epsilon$ of the bolometric luminosity is 
emitted in the X-ray band of $0.3-10\kev$, then using the expression for the 
Eddington luminosity
\begin{equation}
\frac{L_{bol}}{L_{Edd}} = \frac{\dot{M}}{\dot{M}_{Edd}} = \frac{L_X}{\epsilon
  \times 1.28 \times 10^{38} ({\rm~M/M_{\sun}})}
\end{equation}
where $L_X$ is in ${\rm~erg~s^{-1}}$. 
Use of Eq.~\ref{a} with a BB temperature of $kT=170\ev$ for HoII X-1 gives
\begin{equation}
 \frac{\dot{M}}{\dot{M}_{Edd}} \sim  \sqrt{\left[\frac{L_X \times 3.2\times
 10^{-42}}{(\epsilon/0.1)}\right]}
\end{equation}
For $L_X \sim 2\times 10^{40}{\rm~erg~s^{-1}}$ as observed on 10 April 2002,
 $\dot{M}/\dot{M}_{Edd} \sim 0.2$. If we adopt a BB temperature, $kT =
 110\ev$, as suggested from the multi-component model including mekal plasma,
 the relative accretion rate is $\sim 0.1$.

\begin{figure}
 \centering
\includegraphics[width=6cm,angle=-90]{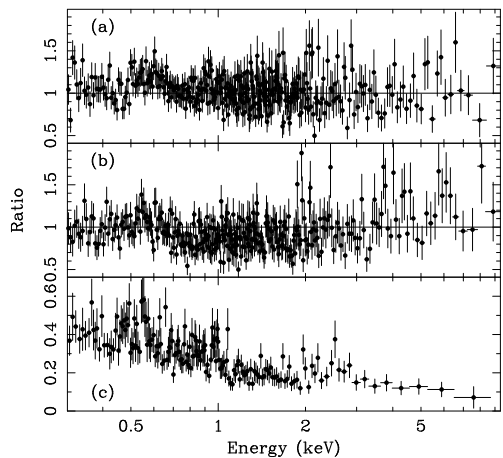}
\caption{A comparison of the observed EPIC PN data of (a) 10 April 2002, (b)
  16 April 2002, and (c) 18 September 2002,  and
  the best-fitting  model of 10 April 2002 consisting of a power law and a
  BB 
  modified by absorption. The ratios are shown  for the normalization of the PN spectrum.  The drop
  in flux below $0.5\kev$ in the spectrum of $10$ April 2002 is present only
  in the PN data and is not seen in any of the MOS spectrum.}
\label{f3}
\end{figure}

\begin{figure}
 \centering
\includegraphics[width=6cm,angle=-90]{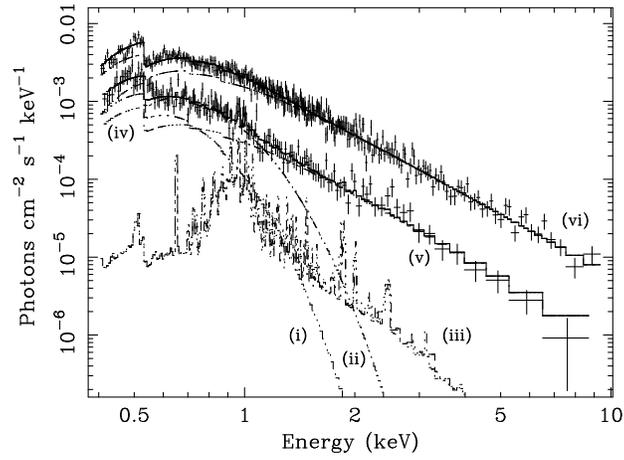}
\caption{A comparison of the unfolded EPIC PN spectra of HoII X-1 in the high (10 April
  2002) and low (18 September 2002) flux states. In both cases, the
  best-fitting model consists of a BB, a power-law, and a mekal. The
  dotted lines are the model components ($i$) BB (low state), ($ii$)
  BB (high state),
  ($iii$) mekal, ($iv$) power law.  The data points
  with the solid line are the observed spectra and the best-fit model for ($v$)
low state and ($vi$) high state. }
\label{f4}
\end{figure}

HoII X-1 resides within a starburst region, therefore it is possible that part
of the X-ray emission arises from the starburst region or a supernova remnant.
Our spectral modeling, after taking into account such a contribution as an
optically-thin emission, suggests that the observed unusual spectral transition to a
low/soft state is due to the drop off the power-law flux alone without any
change in the slope. A higher relative
contribution of the optically-thin emission in the low state resulted in an
overall softer spectrum. Thus,   
we suggest that the X-ray emission from HoII X-1 consists of two distinct
components: ($i$) a soft X-ray excess, described by an MCD
($kT \sim 140\ev$) and a power-law component ($\Gamma \sim2.7$) similar to that
observed from accreting BHs (AGNs and BH binaries), and ($ii$) a thin thermal
component ($kT \sim 0.9\kev$, $L(0.05-10\kev = 3\times
10^{38}{\rm~erg~s^{-1}}$).
In this case, the
spectral variability of the accreting BH in HoII X-1 is not unusual and
similar to many BH X-ray binaries.
 {\it Chandra} ACIS-S observation barely
resolved X-ray emission from HoII X-1, suggesting that the thermal component
arises from a  compact region ($\sim 10{\rm~pc}$; Lehmann et al. 2004).  The estimated luminosity of the thin thermal component
arising from a compact region is not unusually large, as the cooling time,
$\tau_{cool} \sim 50000{\rm~yrs}$ is sufficiently long, where $\tau_{cool}$
was estimated from the temperature, electron density, size, and luminosity of
the mekal plasma.  A compact size and a high
luminosity for the thin thermal component are suggestive of a young supernova
remnant or hypernova remnant. The thin thermal emission explains the line
feature at $\sim 2.5\kev$, detected at $2.8\sigma$ level in the low state
spectrum only, as the S~\small{XV}~K$\alpha$ line (see Figs.~\ref{f3} \& \ref{f4}), however,
there may be some contribution from Si~\small{XIII} RRC.
High resolution X-ray observations will
be required to investigate the nature of the soft component modeled here as thin
thermal emission.

  \acknowledgements 
We thank an anonymous referee and A. R. Rao for fruitful comments and
suggestions. This work is based on observations obtained with \xmm{}, an ESA science 
 mission with
 instruments and contributions directly funded by ESA Member States and
 the USA (NASA). REG acknowledges NASA award NAG5-9902.

\clearpage

\clearpage

\begin{table}
\caption{Best-fit spectral model parameters for Holmberg II X-1.}
\label{tab1}
\begin{tabular}{lcccccc}
\tableline\tableline
   & \multicolumn{2}{c}{10 April 2002} &
\multicolumn{2}{c}{16 April 2002} &  \multicolumn{2}{c}{18 September 2002}  \\
Parameter  & model 1\tablenotemark{a} & model 2\tablenotemark{a} &  model 1\tablenotemark{a} & model 2\tablenotemark{a} &  model 1\tablenotemark{a} & model 2\tablenotemark{a} \\ \tableline
$N_H (10^{21}{\rm~cm^{-2}})$ & $1.6_{-0.1}^{+0.2}$ &$1.6_{-0.1}^{+0.2}$ &
 $1.4_{-0.3}^{+0.3}$  & $1.5_{-0.3}^{+0.3}$ &
$1.4_{-0.4}^{+0.5}$ & $1.5_{-0.3}^{+0.7}$ \\
$kT$(eV) & $141_{-15}^{+18}$ & $187_{-26}^{+28}$ &
$128_{-13}^{+22}$ & $170_{-25}^{+27}$ & $120_{-17}^{+22}$
& $149_{-27}^{+28}$  \\
$f_{bb}$\tablenotemark{b} & $1.9$ & $2.7$ & $2.4$ & $3.4$ & $1.1$ & $1.6$ \\
$\Gamma_X$ & $2.64_{-0.06}^{+0.06}$  & $2.62_{-0.07}^{+0.07}$  &
$2.40_{-0.08}^{+0.07}$ & $2.37_{-0.07}^{+0.08}$ &
  $2.89_{-0.17}^{+0.16}$  & $2.88_{-0.17}^{+0.18}$  \\
$f_{PL}\tablenotemark{c}$ & $10.1$ & $9.7$ & $6.9$ & $6.7$ & $2.4$ & $2.3$  \\
$f_{int}$\tablenotemark{d} & $14.8$  & 15.2 & $12.5$ & $12.9$   & $3.9$ & $4.4$  \\ 
$L_{in}$\tablenotemark{e} & $2.0$ & $2.1$ & $1.7$ & $1.8$ & $0.5$ & $0.6$ \\  
$\chi^2_{min}$/dof & $870.7/756$ & $874.5/756$ & $606.9/562$ & $608.9/562$ &
   $270.0/268$ & $271.7/268$  \\ \tableline 
\end{tabular} 
\tablenotetext{a}{model 1 - simple BB and power-law model modified by
  the Galactic as well intrinsic absorption; model 2 - same as model 1 with
  the simple BB replaced with MCD.}
\tablenotetext{b}{Unabsorbed BB flux in units of
  $10^{-12}{\rm~erg~cm^{-2}~s^{-1}}$ and in the $0.3-2\kev$ band.}
\tablenotetext{c}{Unabsorbed power-law flux in units of
  $10^{-12}{\rm~erg~cm^{-2}~s^{-1}}$  and in the $0.3-2\kev$ band.} 
\tablenotetext{d}{Unabsorbed flux in the $0.3-10\kev$ band and in the units of
  $10^{12}{\rm~erg~cm^{-2}~s^{-1}}$.} 
\tablenotetext{e}{$0.3-10\kev$ intrinsic luminosity in units of
  $10^{40}{\rm~erg~s^{-1}}$  assuming a distance of $3.39{\rm~Mpc}$.}
\end{table}

\end{document}